\documentstyle[11pt,amssymb,amsfonts,amsthm,amsmath,amssymb,amscd,amstext,epsfig]{article}

\def\ben{\begin{equation}}
\def\een{\end{equation}}

\let\w=\omega

\let\pa=\partial
\def\be{\begin{equation}}
\def\ee{\end{equation}}
\def\ba{\begin{array}}
\def\ea{\end{array}}

\def\dalemb#1#2{{\vbox{\hrule height .#2pt
        \hbox{\vrule width.#2pt height#1pt \kern#1pt
                \vrule width.#2pt}
        \hrule height.#2pt}}}

\newcommand{\bea}{\begin{eqnarray}}
\newcommand{\eea}{\end{eqnarray}}
\newcommand{\tr}{{\rm tr} }

\def\Lag{{\mathcal{L}}}

\def\ocal{{\mathcal{O}}}

\begin{document}

\begin{flushright}
NSF-KITP-08-38 \\
PUPT- 2261\\
arXiv:0803.3295 [hep-th]
\end{flushright}

\begin{center}
\vspace{1cm} { \LARGE {\bf Building an AdS/CFT superconductor}}

\vspace{1.1cm}

Sean A. Hartnoll$^\flat$, Christopher P. Herzog$^\sharp$ and Gary
T. Horowitz$^\natural$

\vspace{0.7cm}

{\it $^\flat$ KITP, University of California\\
     Santa Barbara, CA 93106, USA }

\vspace{0.7cm}

{\it $^\sharp$ Department of Physics, Princeton University \\
     Princeton, NJ 08544, USA }

\vspace{0.7cm}

{\it $^\natural$ Department of Physics, UCSB \\
     Santa Barbara, CA 93106, USA }

\vspace{0.7cm}

{\tt hartnoll@kitp.ucsb.edu, cpherzog@princeton.edu, gary@physics.ucsb.edu} \\

\vspace{1.5cm}

\end{center}

\begin{abstract}
\noindent
We show that a simple gravitational theory can provide a
holographically dual description of a superconductor. There is a
critical temperature, below which a charged condensate forms via a
second order phase transition and the (DC) conductivity becomes
infinite. The frequency dependent conductivity develops a gap
determined by the condensate. We find evidence that the condensate consists
of pairs of quasiparticles.

\end{abstract}

\pagebreak
\setcounter{page}{1}

\section{Introduction}

A most remarkable result to emerge from string theory is
the AdS/CFT correspondence \cite{Maldacena:1997re}, which relates
string theory on asymptotically anti de Sitter spacetimes to a
conformal field theory on the boundary. In recent years, it has
become clear that this holographic correspondence between a
gravitational theory and a quantum field theory can be
extended to describe aspects
of nuclear physics such as the results of heavy ion collisions at
RHIC \cite{Mateos:2007ay} and to certain
condensed matter systems. Phenomena such as the Hall
effect \cite{Hartnoll:2007ai} and Nernst effect
\cite{Hartnoll:2007ih, Hartnoll:2007ip, Hartnoll:2008hs}
have dual gravitational
descriptions. One can ask if there is a dual
gravitational description of superconductivity.

Conventional superconductors, including many metallic elements
(Al, Nb, Pb, ...), are well described by BCS theory \cite{parks}.
However, basic aspects of unconventional superconductors,
including the pairing mechanism, remain incompletely understood.
There are many indications that the normal state in these
materials is not described by the standard Fermi liquid theory
\cite{hightc}. We therefore hope that a tractable theoretical model of a
strongly coupled system which develops superconductivity will be
of interest. Several important unconventional superconductors,
such as the cuprates and organics, are layered and much of the
physics is 2+1 dimensional. Our model will also be 2+1
dimensional.

To map a superconductor to a gravity dual, we introduce
temperature by adding a black hole \cite{Witten:1998qj} and a
condensate through a charged scalar field. To reproduce the
superconductor phase diagram, we require a system that admits
black holes with scalar hair at low temperature, but no hair at
high temperature. While Hertog has shown that neutral AdS black
holes can have neutral scalar hair only if the theory is unstable
\cite{Hertog:2006rr}, Gubser has recently suggested that a charged
black hole will support  charged scalar hair if the charges are
large enough \cite{Gubser:2008px}. We consider a simpler version
of Gubser's bulk theory (in which the black hole can remain
neutral) and show that it indeed provides a dual description of a
superconductor.

\section{The model: condensing charged operators}

We start with the  planar Schwarzschild anti-de Sitter black hole
\be\label{sads}
ds^2 = - f(r) dt^2 + \frac{dr^2}{f(r)} + r^2 (dx^2 + dy^2) \,,
\ee
where
\be
f = \frac{r^2}{L^2} - \frac{M}{r} \,.
\ee
$L$ is the AdS radius and $M$ determines the Hawking temperature of the black hole:
\be
T = \frac{3 M^{1/3}}{4 \pi L^{4/3}} \,.
\ee
This black hole is 3+1 dimensional, and so will be dual to a 2+1
dimensional theory. In this background, we now consider a Maxwell
field and a charged complex scalar field, with Lagrangian density\footnote{%
Introducing a gauge coupling $1/e^2$ in front of the $|F|^2$ term in
the action is equivalent to rescaling the fields $\Psi \to e \Psi$
and $A_\mu \to e A_\mu$.  Setting $e=1$ is a choice of units of charge
in the dual 2+1 theory.
}
\be\label{eq:lag}
\Lag = - \frac{1}{4} F^{ab} F_{ab} - V(|\Psi|) - |\pa \Psi - i A \Psi |^2 \,.
\ee
For simplicity and concreteness, we will focus on the case
\be\label{eq:mass}
V(|\Psi|) = - \frac{2 |\Psi|^2}{L^2} \,.
\ee
Although the mass squared is negative, it is above
the Breitenlohner-Freedman bound \cite{Breitenlohner:1982jf} and
hence does not induce an instability. It corresponds to a
conformally coupled scalar in our background (\ref{sads}) and
 arises in several contexts in which the $AdS_4/CFT_3$
correspondence is embedded into string theory. For instance, the
truncation of M theory on $AdS_4 \times S^7$ to ${\mathcal{N}} =
8$ gauged supergravity has scalars and pseudoscalars with this
mass, dual to the bilinear operators $\tr \Phi^{(I} \Phi^{J)}$ and
$\tr \Psi^{(I} \Psi^{J)}$ in the dual ${\mathcal{N}} = 8$ Super
Yang-Mills theory, respectively. However, we should note that our
Lagrangian (\ref{eq:lag}) has not been obtained from M theory. We expect that our choice of mass is not crucial, and qualitatively similar results will hold, e.g., for massless fields.

We will work in a limit in which the Maxwell field and scalar
field do not backreact on the metric. This limit is consistent
as long as the fields are small in Planck units. (Recall that in
the analogous case one dimension higher, only Yang-Mills states
with energy of order $N^2$ have finite backreaction in the bulk.)
Alternatively, this decoupled Abelian-Higgs sector can be obtained from the
full Einstein-Maxwell-scalar theory considered in \cite{Gubser:2008px}
through a scaling limit in which the product
of the charge of the black hole and the charge of the scalar field is held
fixed while the latter is taken to infinity.  Thus we will obtain solutions of non-backreacting scalar hair on
the black hole.
As we shall see, our simple
model captures the physics of interest.

Taking a plane symmetric ansatz, $\Psi = \Psi(r)$, the scalar
field equation of motion is
\be\label{eq:Psieq}
\Psi'' + \left(\frac{f'}{f} + \frac{2}{r}\right) \Psi' +
\frac{\Phi^2}{f^2}\Psi + \frac{2}{L^2 f} \Psi = 0 \,,
\ee
where the scalar potential $A_t = \Phi$. With $A_r = A_x = A_y =
0$, the Maxwell equations imply that the phase of $\Psi$ must be
constant. Without loss of generality we therefore take $\Psi$ to
be real. The equation for the scalar potential $\Phi$ is the time
component of the equation of motion for a massive vector field
\be\label{eq:Phieq}
\Phi'' + \frac{2}{r} \Phi' - \frac{2 \Psi^2}{f} \Phi = 0 \,,
\ee
where $2 \Psi^2$ is the, in our case, $r$ dependent mass. The
charged condensate has triggered a Higgs mechanism in the bulk
theory. At the horizon, $r=r_0$, for $\Phi
\, dt$ to have finite norm, $\Phi = 0$, and (\ref{eq:Psieq}) then
implies $\Psi = - 3r_0
\Psi'/2$. Thus, there is a two parameter family of solutions which
are regular at the horizon. Integrating out to infinity, these
solutions behave as
\be
\Psi = \frac{\Psi^{(1)}}{r} + \frac{\Psi^{(2)}}{r^2} + \cdots \,.
\ee
and
\be\label{asympphi}
\Phi = \mu - \frac{\rho}{r} + \cdots \,.
\ee
For $\Psi$, both of these falloffs are normalizable
\cite{Klebanov:1999tb}, so one can impose the boundary condition
that either one vanishes.\footnote{%
 One might also imagine imposing boundary
 conditions in which both $\Psi^{(1)}$ and $\Psi^{(2)}$ are
 nonzero. However, if these boundary conditions respect
 the AdS symmetries, then the
 result is a theory in which the asymptotic AdS region is
 unstable \cite{Hertog:2004rz}.
}
After imposing the
condition that either $\Psi^{(1)}$ or $\Psi^{(2)}$ vanish we have
a one parameter family of solutions.

It follows from (\ref{eq:Phieq}) that the solution for $\Phi$ is
always monotonic: It starts at zero and cannot have a positive
maximum or a negative minimum. Note that even though the field
equations are nonlinear, the overall signs of $\Phi$ and $\Psi$
are not fixed. We will take $\Phi$ to be positive and hence have a system
with positive charge density. The sign of $\Psi$ is
part of the freedom to choose the overall phase of $\Psi$.

Properties of the dual field theory can be read off from the
asymptotic behavior of the solution. For example, the asymptotic
behavior (\ref{asympphi}) of $\Phi$ yields the chemical potential
$\mu$ and charge density $\rho$ of the field theory. The
condensate of the scalar operator $\ocal$ in the field theory dual
to the field $\Psi$  is given by
\be
\langle \ocal_i\rangle = \sqrt{2} \Psi^{(i)} \,,\quad i=1,2
\ee
with the boundary condition $\epsilon_{ij} \Psi^{(j)} = 0$. The
$\sqrt{2}$ normalization simplifies subsequent formulae, and corresponds
to taking the bulk-boundary coupling $\frac{1}{2} \int d^3x \left(\bar \ocal \Psi + \ocal \bar
\Psi \right)$. Note that $\ocal_i$ is an operator with dimension $i$.
From this point on we will work in units in which the AdS radius
is $L=1$. Recall that $T$ has mass dimension one, and $\rho$ has
mass dimension two so $\langle
\ocal_i\rangle/T^i$ and $\rho/T^2$ are dimensionless quantities.

\begin{figure}[h]
\begin{center}
\epsfig{file=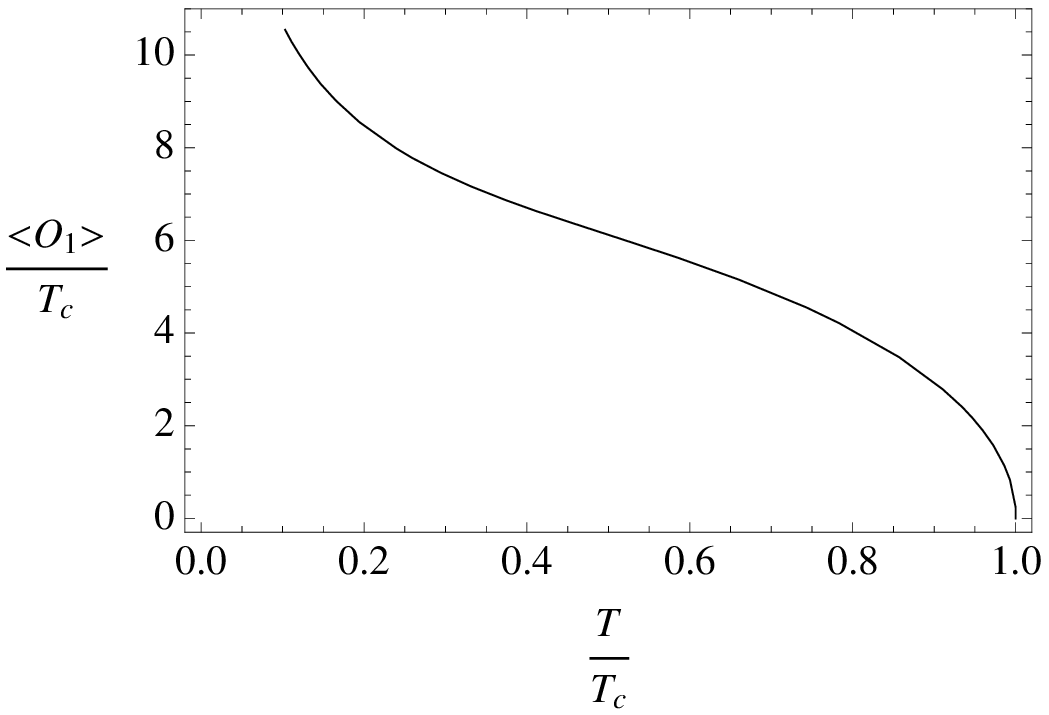,width=2.9in,angle=0,trim=0 0 0 0}%
\hspace{0.2cm}\epsfig{file=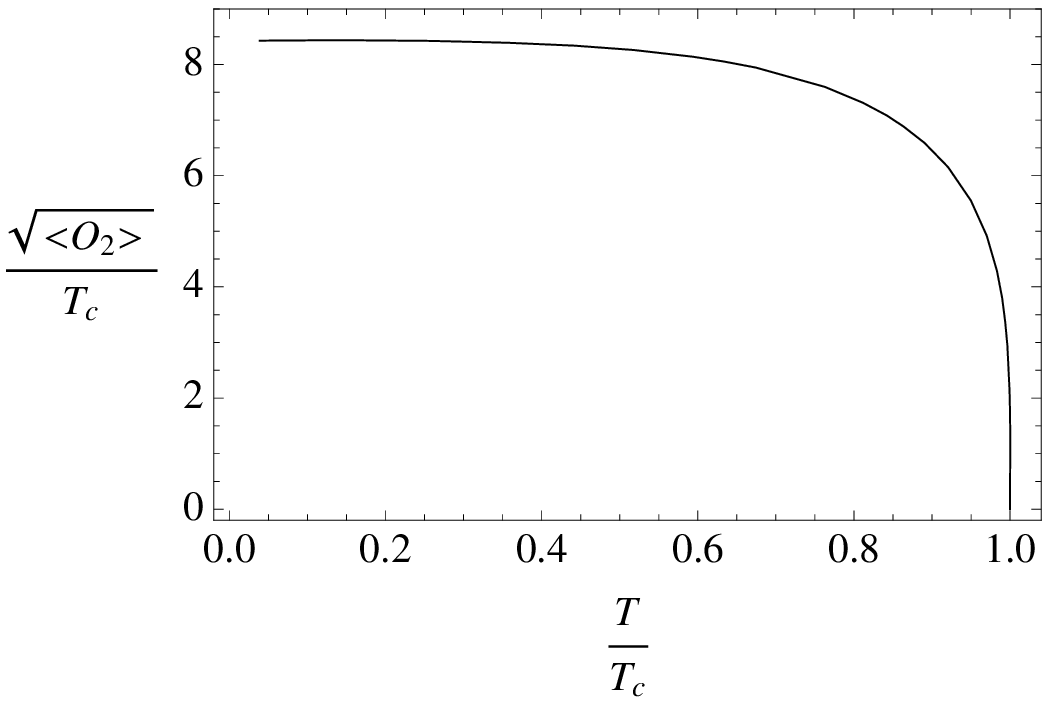,width=2.9in,angle=0,trim=0 0 0 0}%
\end{center}
\caption{The condensate as a function of
temperature for the two operators $\ocal_1$ and $\ocal_2$. The
condensate goes to zero at $T=T_c \propto \rho^{1/2}$.
  \label{fig:condensate}}
\end{figure}

An exact solution to eqs (\ref{eq:Psieq},\ref{eq:Phieq}) is
clearly $\Psi=0$ and $\Phi = \mu - \rho/r$. It appears difficult
to find other analytic solutions to these nonlinear equations.
However, it is straightforward to solve them numerically. We  find
that solutions exist with all values of the condensate $\langle
\ocal \rangle$. However, as shown in figure \ref{fig:condensate},
in order for the operator to condense, a minimal ratio of charge
density over temperature squared is required.

The right hand curve in figure 1 is qualitatively similar to that
obtained in BCS theory, and observed in many materials, where the
condensate goes to a constant at zero temperature. The left hand
curve starts similarly, but at low temperature the condensate
appears to diverge as $T^{-1/6}$. However, when the condensate becomes
very large, the backreaction on the bulk metric can no longer be neglected.
At extremely low temperatures, we will eventually be outside
the region of validity of our approximation.

By fitting these curves, we see that for small condensate there is
a square root behaviour that is typical of second order phase
transitions. Specifically, for one boundary condition we find
\be
\langle \ocal_1 \rangle \approx \, 9.3 \, T_c \, (1-T/T_c)^{1/2} \,,
\quad \text{as} \quad T \to T_c \,,
\ee
where the critical temperature is $T_c \approx 0.226
\rho^{1/2}$. For the other boundary condition
\be
\langle \ocal_2 \rangle \approx 144 \, T_c^2 \, (1 - T/T_c)^{1/2}
\,, \quad \text{as} \quad T \to T_c \,,
\ee
where now $T_c \approx 0.118 \rho^{1/2}$. The continuity of the
transition can be checked by computing the free energy. Finite
temperature continuous symmetry breaking phase transitions are
only possible in 2+1 dimensions in the large $N$ limit (i.e.\ the
classical gravity limit of our model), where fluctuations are
suppressed. These transitions will become crossovers at finite
$N$.

Thus for $T < T_c$ a charged scalar operator, $\langle \ocal_1
\rangle$ or $\langle \ocal_2
\rangle$, has condensed. It is natural to expect that this
condensate will lead to superconductivity of the current
associated with this charge.

\section{Maxwell perturbations and the conductivity}

We now compute the conductivity in the dual CFT as a
function of frequency. As a first step, we need to solve for
fluctuations of the vector potential $A_x$ in the bulk.  The
Maxwell equation at zero spatial momentum and with a time
dependence of the form $e^{- i \w t}$ gives
\be\label{eq:Axeq}
A_x'' + \frac{f'}{f} A_x' + \left(\frac{\w^2}{f^2} - \frac{2
\Psi^2}{f} \right) A_x = 0 \,.
\ee
To compute causal behavior,
we solve this equation with ingoing wave boundary conditions at the horizon
\cite{Son:2002sd}:  $A_x  \propto f^{-i\omega/3r_0}$.
The asymptotic behaviour of the Maxwell field at large radius is
seen to be
\be
A_x = A_x^{(0)} + \frac{A_x^{(1)}}{r} + \cdots
\ee
The AdS/CFT dictionary tells us that the dual source and
expectation value for the current are given by
\be
A_x = A_x^{(0)} \,, \qquad \langle J_x \rangle = A_x^{(1)} \,.
\ee
Now from Ohm's law we can obtain the conductivity
\be\label{eq:conductivity}
\sigma(\w) = \frac{\langle J_x \rangle}{E_x} = - \frac{ \langle J_x \rangle}{\dot A_x} = -\frac{ i \langle J_x \rangle}{\w
A_x} = - \frac{i A_x^{(1)}}{\w A_x^{(0)}} \,.
\ee

\begin{figure}[h]
\begin{center}
\epsfig{file=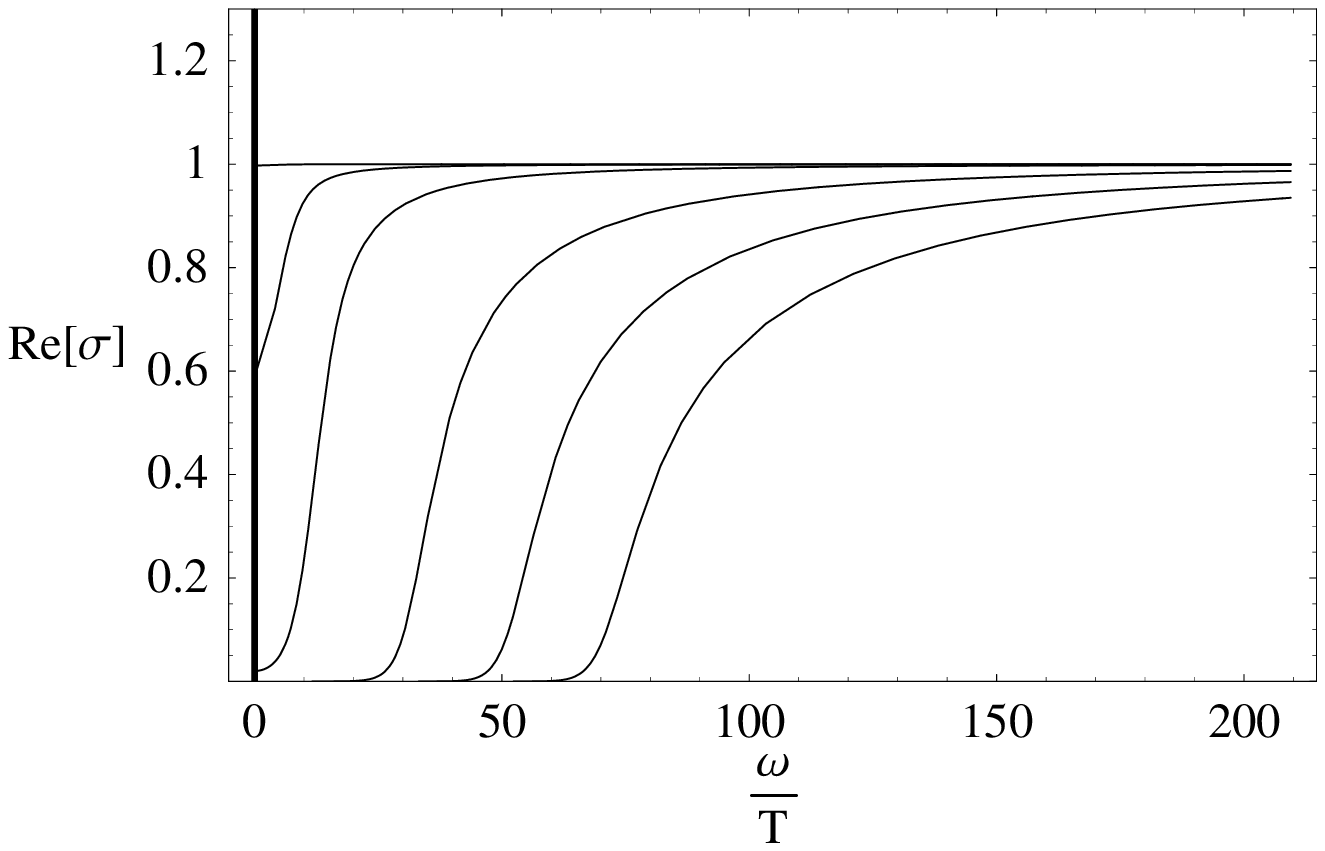,width=2.9in,angle=0,trim=0 0 0 0}%
\hspace{0.2cm}\epsfig{file=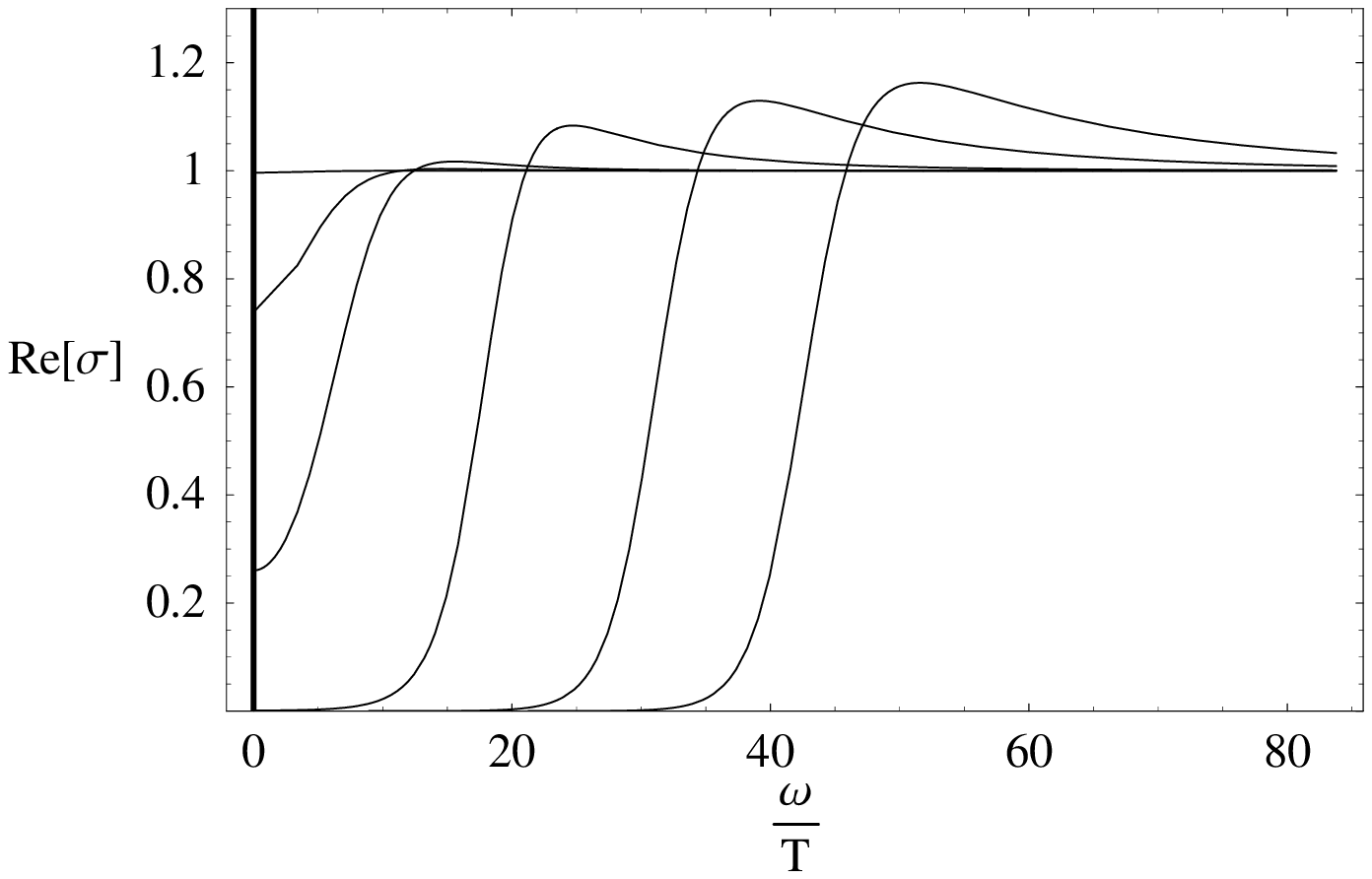,width=2.9in,angle=0,trim=0 0 0 0}%
\end{center}
\caption{The formation of a gap in the real, dissipative,
part of the conductivity as the temperature is lowered below the
critical temperature. Results shown for both the $\ocal_1$
operator (left) and the $\ocal_2$ operator (right). There is also a
delta function at $\w = 0$. The rightmost curve in each plot
corresponds to $T/T_c = .0066$ (left) and
$T/T_c = .0026$ (right).
  \label{fig:gap}}
\end{figure}

In figure \ref{fig:gap} we plot the frequency dependent
conductivity obtained by solving (\ref{eq:Axeq}) numerically. The
horizontal line corresponds to temperatures at or above the
critical value, where there is no condensate. The fact that the
conductivity in the normal phase is frequency independent is a
characteristic of theories with $AdS_4$ duals \cite{Herzog:2007ij}.
The subsequent curves describe successively lower values of the
temperature (for fixed charge density).  We see that as the
temperature is lowered, a gap opens. The gap becomes increasingly
deep until the (real part of the) conductivity is exponentially
small.

There is also a delta function at $\omega=0$ which appears as soon
as $T < T_c$. This can be seen by looking at the imaginary part of
the conductivity. The Kramers-Kronig relations relate the real and
imaginary parts of any causal quantity, such as the conductivity,
when expressed in frequency space. They may be derived from
contour integration and the fact that functions that vanish when
$t<0$, when transformed to frequency space only have poles in the
lower half plane (in our conventions). One of the relations is
\be\label{eq:kramers}
\text{Im} [\sigma(\w)] = - \frac{1}{\pi} {\mathcal{P}}
\int_{-\infty}^{\infty} \frac{\text{Re} [\sigma(\w')] d\w'}{\w'-\w}
\,.
\ee
From this formula we can see that  the real part of the
conductivity contains a delta function, $\text{Re} [\sigma(\w)] =
\pi \delta(\w)$, if and only if the imaginary part has a pole,
$\text{Im} [\sigma(\w)] = 1/\w$. One finds that there is indeed a
pole in Im$[\sigma]$ at $\omega=0$ for all $T<T_c$. The superfluid
density is the coefficient of the delta function of the real part
of the conductivity\footnote{The superfluid density is usually
defined as the coefficient of $\delta(\w)$ multiplied by the mass
of the electron. In simple two fluid models, this density is
related to the London magnetic penetration depth, $n_s = 1/ 4 \pi
\lambda_L^2$. Our scaling (\ref{eq:ns}) thus implies $\lambda_L \sim (T_c-T)^{-1/2}$,
consistent with Landau-Ginzburg theory.}
\be
\text{Re}[\sigma(\w)] \sim \pi n_s \delta(\w) \,.
\ee
By (\ref{eq:kramers}), $n_s$ is also the coefficient of the pole
in the imaginary part $\text{Im}[\sigma(\w)] \sim n_s/\w$ as $\w
\to 0$. We find that the superfluid density vanishes linearly with
$T_c-T$:
\be\label{eq:ns}
n_s \approx C_i \, (T_c - T) \,  \quad \text{as} \quad T \to T_c \
,
\ee
where $C_1 = 16.5$ for the $\ocal_1$ theory while $C_2 = 24$ for
the $\ocal_2$ theory.

\begin{figure}[h]
\begin{center}
\epsfig{file=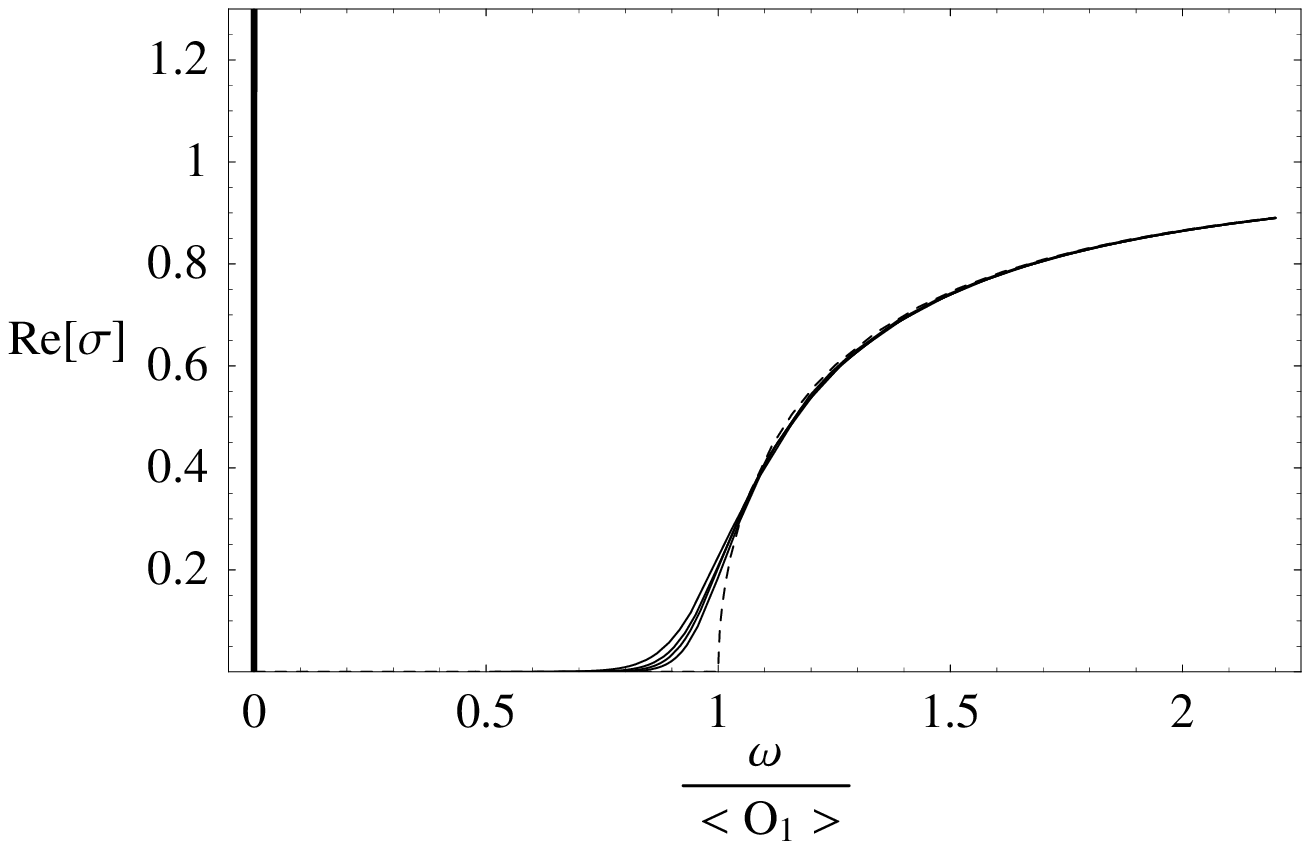,width=2.9in,angle=0,trim=0 0 0 0}%
\hspace{0.2cm}\epsfig{file=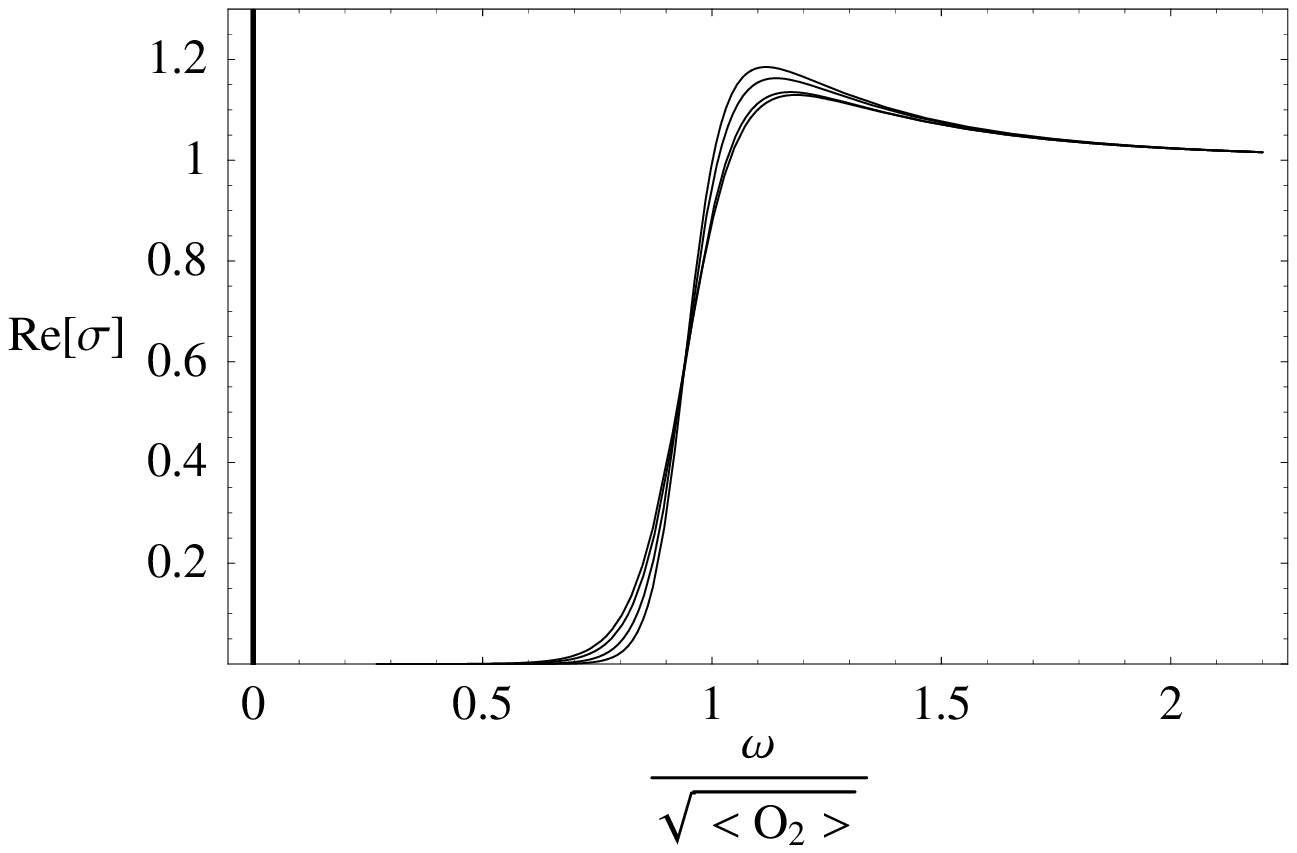,width=2.9in,angle=0,trim=0 0 0 0}%
\end{center}
\caption{The gap at small $T/T_c$, with the frequency normalised in terms of the
condensate. On the right the gap is finite, but since $\langle
\ocal_1 \rangle$ becomes large at small $T$, the gap on the left
is also becoming large. The dashed curve on the left plot is
(\ref{eq:sqrtlaw}).
  \label{fig:gap2}}
\end{figure}

In figure \ref{fig:gap2} we rescaled the small $T/T_c$
plots of figure \ref{fig:gap} by plotting the frequency in units
of the condensate rather than the temperature.
The curves tend to a limit in which the width of
the gap is proportional to the size of the condensate.
The differing shapes of the plots in figure \ref{fig:gap2}
are precisely what is expected from type II and type I coherence
factors, respectively \cite{parks}. Type II coherence suppresses
absorption near the edge of the gap, explaining the slower
rise of $\mbox{Re}[\sigma] $ in the left hand plot. It is
possible that this difference is due to the operator $\ocal_1$
being a pair of bosons and $\ocal_2$ a pair of fermions,
as in the case of $AdS_4 \times S^7$.

The Ferrell-Glover sum rule states that
$\int \mbox{Re}[\sigma]  d\omega$ is a constant independent of
temperature.  Thus the area missing under the curve
$\mbox{Re}[\sigma]$ due to the gap must be made up
by the delta function at $\omega =0$.  That $\mbox{Re}[\sigma]$
exceeds the value one in figure \ref{fig:gap2} (right) implies
then that the superfluid density $n_s$ must be correspondingly reduced
for the $\ocal_2$ system compared with the $\ocal_1$ system for $T \ll T_c$.

We can also compute the contribution of the normal, non-superconducting,
component to the DC conductivity. Let us define
\be
n_n = \lim_{\w \to 0} \text{Re} [\sigma(\w)] \,.
\ee
From our numerics we obtain
\be\label{eq:nn}
n_n \sim e^{- \Delta/T} \,, \quad \text{for} \quad
\frac{\Delta}{T} \gg 1 \,,
\ee
where we have $\Delta = \langle \ocal_1 \rangle/2$ and $
\Delta = \sqrt{\langle \ocal_2 \rangle}/2$. Numerically this factor of $1/2$
is accurate to at least $4\%$. From (\ref{eq:nn}), $\Delta$ is
immediately interpreted as the energy gap for charged excitations.
The gap we found previously in the frequency dependent
conductivity was $2
\Delta$. The extra factor of two is expected if the gapped charged
quasiparticles are produced in pairs, suggesting that there is a
`pairing mechanism' at work in our model. Our results for $\Delta$
are suggestive of strong pairing interactions. Note that for
figure
\ref{fig:condensate} (right) at $T=0$ we find $2 \Delta \approx
8.4 T_c$, which might be compared with the BCS prediction $2
\Delta \approx 3.54 T_c$. The larger value is what one expects for
deeply bound Cooper pairs. Indeed, in our other model, figure
\ref{fig:condensate} (left), we see that $\Delta$ actually
diverges at low $T$.

\begin{figure}[h]
\begin{center}
\epsfig{file=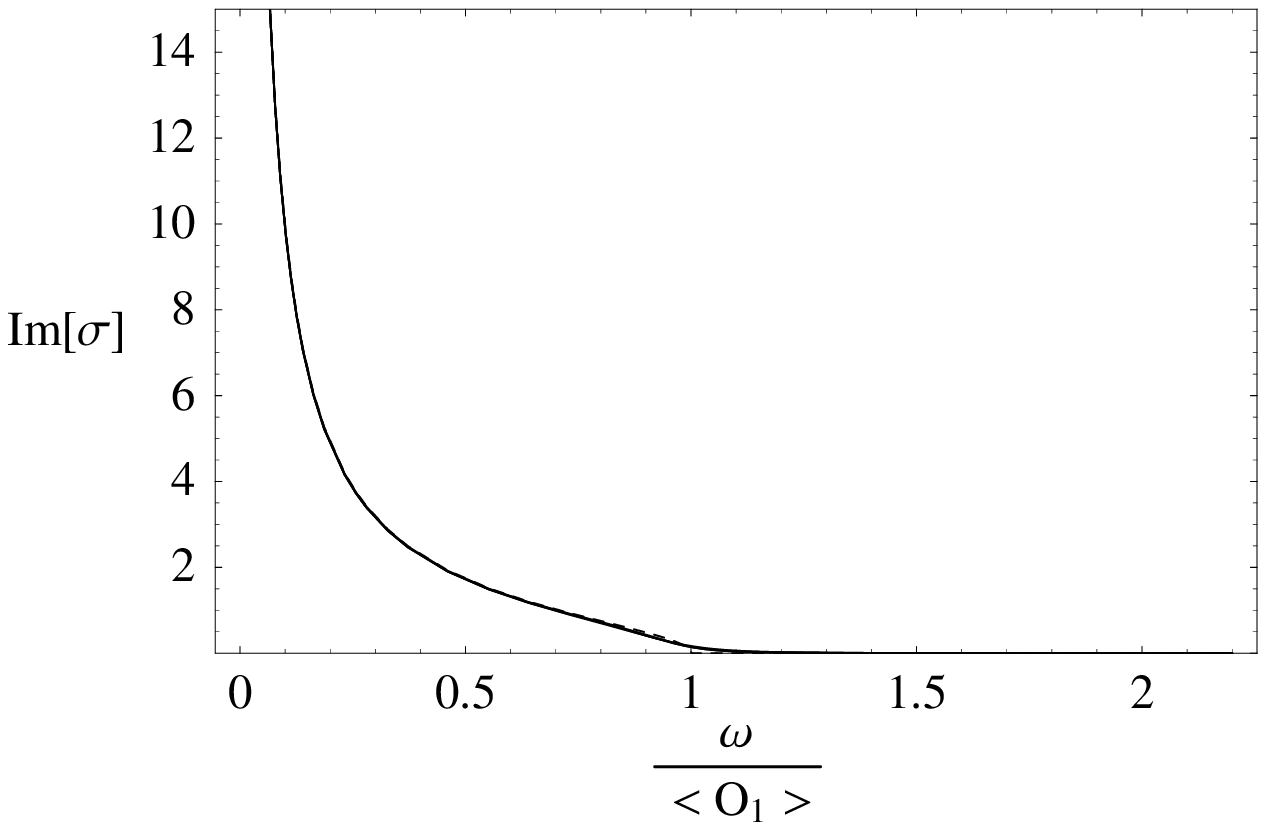,width=2.9in,angle=0,trim=0 0 0 0}%
\hspace{0.2cm}\epsfig{file=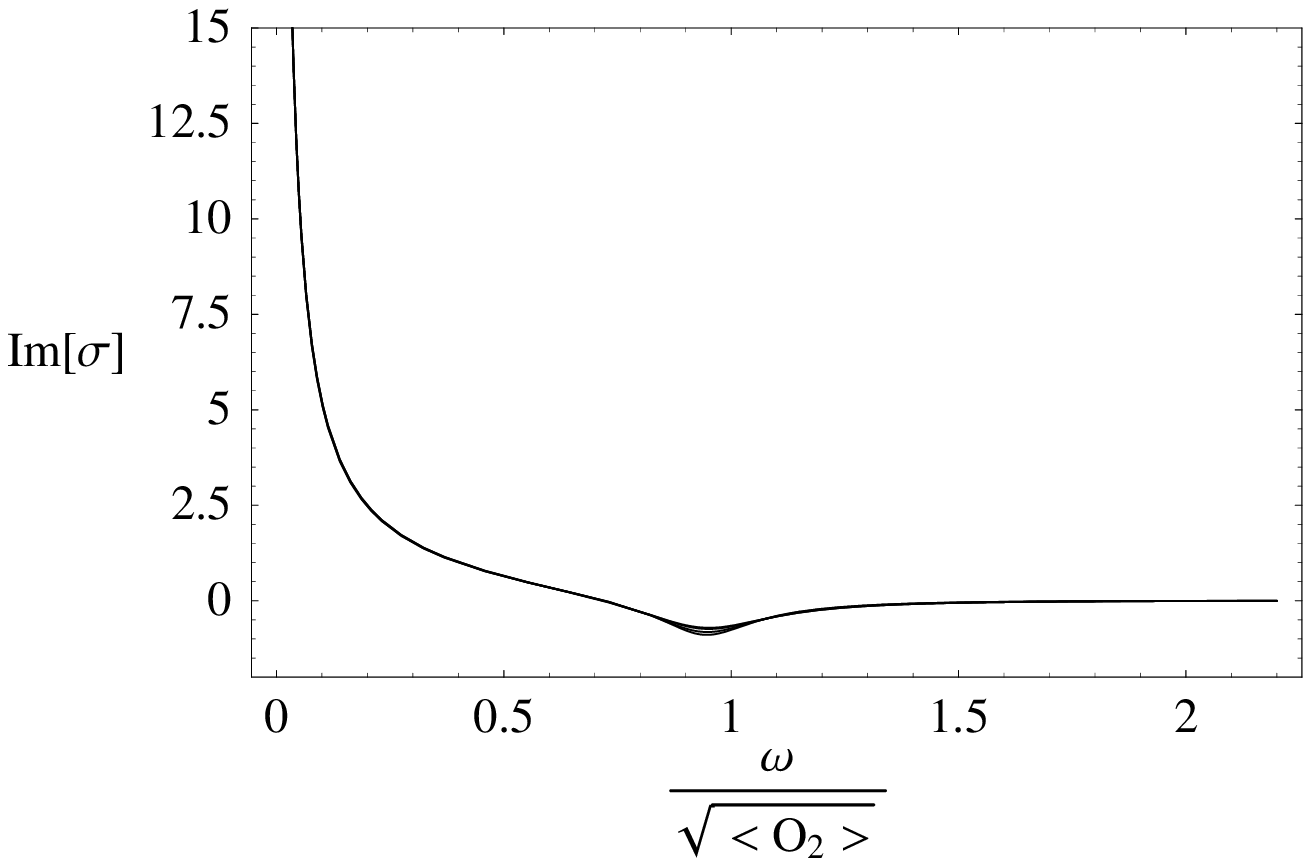,width=2.9in,angle=0,trim=0 0 0 0}%
\end{center}
\caption{The imaginary part of the conductivity at small $T/T_c$,
with the frequency normalised in terms of the condensate. The
analytic expression (\ref{eq:sqrtlaw}) is also shown on the left, but is
indistinguishable from the numerics.
  \label{fig:gap3}}
\end{figure}

Finally in figure \ref{fig:gap3} we plot the imaginary part of the
conductivity in this limiting, low temperature, regime. Again we
see that the curves rapidly approach a limiting curve. We also see
the advertised pole at $\w=0$.

It is natural to ask if one can reproduce this limiting low
temperature behavior by just taking $M=0$ in our background
metric, and considering our matter fields in anti de Sitter space.
One problem is that there are no solutions to the field equations
(\ref{eq:Psieq},\ref{eq:Phieq}) which are smooth on the horizon of
the Poincare patch. Nevertheless, for the $\ocal_1$ case, we have
observed numerically that at low temperatures, $\Psi \approx
\langle  \ocal_1 \rangle / \sqrt{2} r$. Taking $M\to 0$ where $f
\approx r^2$, (\ref{eq:Axeq}) can be solved exactly to yield $A_x
= A_{x}^{(0)} \exp(\pm \sqrt{\langle \ocal_1 \rangle^2 -
\omega^2}/r)$. This exact result then produces the nonzero conductivities
\be\label{eq:sqrtlaw}
\text{Re}[\sigma] = \frac{\sqrt{\w^2 - \langle \ocal_1 \rangle^2}}{\omega} \quad \text{for}
\quad \w > \langle \ocal_1 \rangle \ , \quad
\text{Im}[\sigma] = \frac{\sqrt{\langle \ocal_1 \rangle^2 - \w^2}}{\omega} \quad \text{for}
\quad \w < \langle \ocal_1 \rangle \ ,
\ee
via (\ref{eq:conductivity}). The curves on the left hand sides of
figures \ref{fig:gap2} and \ref{fig:gap3} are well approximated by
the conductivity (\ref{eq:sqrtlaw}). We have included
(\ref{eq:sqrtlaw}) as a dashed curve in these plots.

\section{Discussion}

We have shown that a simple 3+1 dimensional bulk theory can reproduce
several properties of a 2+1 dimensional superconductor. Below
a second order superconducting phase transition the DC superconductivity
becomes infinite and an energy gap for charged excitations is formed.

There are many extensions of this model that we hope to consider
elsewhere:
1) By probing the system with spatially varying fields and an
external magnetic field, one can compute the superconducting coherence
length and penetration depth, respectively.
2) One would like to consider a wider class of models by
allowing for more general masses for the charged scalar field.
3) One should
study the effects of backreaction on the bulk spacetime metric.
4) Perhaps the most interesting question is to understand the
`pairing mechanism' in field theory that leads to a condensate in
these systems.

\subsection*{Acknowledgements}
We would like to thank A.~Bernevig, S.~Gubser, D.~Huse, P.~Kovtun and D.~Mateos for
discussion. This work was supported in part by NSF grants
PHY-0243680, PHY-0555669 and PHY05-51164.


\begin{thebibliography}{99}

\bibitem{Maldacena:1997re}
J.~M.~Maldacena, ``The large N limit of superconformal field
theories and supergravity,'' Adv.\ Theor.\ Math.\ Phys.\  {\bf 2}
(1998) 231 [Int.\ J.\ Theor.\ Phys.\  {\bf 38} (1999) 1113]
[arXiv:hep-th/9711200].

\bibitem{Mateos:2007ay}
  D.~Mateos,
  ``String Theory and Quantum Chromodynamics,''
  Class.\ Quant.\ Grav.\  {\bf 24}, S713 (2007)
  [arXiv:0709.1523 [hep-th]].

\bibitem{Hartnoll:2007ai}
  S.~A.~Hartnoll and P.~Kovtun,
  ``Hall conductivity from dyonic black holes,''
  Phys.\ Rev.\  D {\bf 76}, 066001 (2007)
  [arXiv:0704.1160 [hep-th]].

\bibitem{Hartnoll:2007ih}
  S.~A.~Hartnoll, P.~K.~Kovtun, M.~Muller and S.~Sachdev,
  ``Theory of the Nernst effect near quantum phase transitions in condensed
  matter, and in dyonic black holes,''
  Phys.\ Rev.\  B {\bf 76}, 144502 (2007)
  [arXiv:0706.3215 [cond-mat.str-el]].

\bibitem{Hartnoll:2007ip}
  S.~A.~Hartnoll and C.~P.~Herzog,
  ``Ohm's Law at strong coupling: S duality and the cyclotron resonance,''
  Phys.\ Rev.\  D {\bf 76}, 106012 (2007)
  [arXiv:0706.3228 [hep-th]].

\bibitem{Hartnoll:2008hs}
  S.~A.~Hartnoll and C.~P.~Herzog,
  ``Impure AdS/CFT,''
  arXiv:0801.1693 [hep-th].

\bibitem{parks}
  R.~D.~Parks, \textit{Superconductivity}, Marcel Dekker Inc.
  (1969).

\bibitem{hightc}
  E.~W.~Carlson, V.~J.~Emery, S.~A.~Kivelson and D.~Orgad,
  ``Concepts in high temperature superconductivity,''
  arXiv:cond-mat/0206217.

\bibitem{Witten:1998qj}
  E.~Witten,
  ``Anti-de Sitter space and holography,''
  Adv.\ Theor.\ Math.\ Phys.\  {\bf 2}, 253 (1998)
  [arXiv:hep-th/9802150].

\bibitem{Hertog:2006rr}
  T.~Hertog,
  ``Towards a novel no-hair theorem for black holes,''
  Phys.\ Rev.\  D {\bf 74}, 084008 (2006)
  [arXiv:gr-qc/0608075].

\bibitem{Gubser:2008px}
  S.~S.~Gubser,
  ``Breaking an Abelian gauge symmetry near a black hole horizon,''
  arXiv:0801.2977 [hep-th].

\bibitem{Breitenlohner:1982jf}
  P.~Breitenlohner and D.~Z.~Freedman,
  ``Stability In Gauged Extended Supergravity,''
  Annals Phys.\  {\bf 144}, 249 (1982).

\bibitem{Klebanov:1999tb}
  I.~R.~Klebanov and E.~Witten,
  ``AdS/CFT correspondence and symmetry breaking,''
  Nucl.\ Phys.\  B {\bf 556}, 89 (1999)
  [arXiv:hep-th/9905104].

\bibitem{Hertog:2004rz}
  T.~Hertog and G.~T.~Horowitz,
  ``Towards a big crunch dual,''
  JHEP {\bf 0407}, 073 (2004)
  [arXiv:hep-th/0406134];
  T.~Hertog and G.~T.~Horowitz,
  ``Holographic description of AdS cosmologies,''
  JHEP {\bf 0504}, 005 (2005)
  [arXiv:hep-th/0503071].

\bibitem{Herzog:2007ij}
  C.~P.~Herzog, P.~Kovtun, S.~Sachdev and D.~T.~Son,
  ``Quantum critical transport, duality, and M-theory,''
  Phys.\ Rev.\  D {\bf 75}, 085020 (2007)
  [arXiv:hep-th/0701036].


\bibitem{Son:2002sd}
  D.~T.~Son and A.~O.~Starinets,
  ``Minkowski-space correlators in AdS/CFT correspondence: Recipe and
  applications,''
  JHEP {\bf 0209}, 042 (2002)
  [arXiv:hep-th/0205051].


\end{thebibliography}
\end{document}